\documentclass[10pt,twoside,BCOR7mm,DIV15,headinclude,footexclude,
               cleardoubleempty,idxtotoc]
{scrartcl}

\usepackage{natbib}
\usepackage[font=small,labelfont=bf]{caption}
\usepackage[english]{babel}
\usepackage{graphicx}
\usepackage{hyperref}
\usepackage{scrpage2}
\usepackage{ifthen}
\usepackage{booktabs}
\usepackage{amsmath}
\usepackage{amssymb}
\usepackage{multicol}
\usepackage{float}
\usepackage{hyperref}

\hypersetup{breaklinks=true
,colorlinks=true,linkcolor=black,urlcolor=blue
,citecolor=black}

\addto\captionsenglish{%
}

\makeatletter
\renewcommand{\@biblabel}[1]{}
\renewcommand{\@cite}[2]{%
{#1\ifthenelse{\boolean{@tempswa}}{,#2}{}}}
\makeatother
\ofoot{\thepage}
\ifoot{}

\setcapindent{0em}
\setheadsepline{1pt}

\setkomafont{pagehead}{\normalfont\sffamily}
\setkomafont{pagenumber}{\normalfont\rmfamily}

\makeatletter
\newcommand{\listofcontributions}{\@starttoc{con}}

\newcommand{\l@contribution} {\@dottedtocline{1}{1.5em}{2.3em}}
\makeatother

\newenvironment{contribution}{
\setcounter{section}{0}
\setcounter{figure}{0}
\setcounter{table}{0}
}{
\newpage
\lehead{}
\rohead{}
}

\def\apj{ApJ}%
\def\aap{A\&A}%
\def\mnras{MNRAS}%
\begin{document}

\setlength{\baselineskip}{2.5ex}

\begin{contribution}
% EXAMPLE AND TEMPLATE FILE FOR PROCEEDINGS OF THE WOLF-RAYET WORKSHOP.
% PLEASE REPLACE THE TEMPLATE TEXT BY YOUR OWN ARTICLE.
% NOTE THAT YOU MUST NOT PROCESS THIS FILE, BUT THE MASTER FILE:
% latex masterfile; dvips masterfile

% RUNNING AUTHOR: PUT AUTHOR NAMED HERE
\lehead{A. Mesa-Delgado, C. Esteban \& J. Garc\'{\i}a-Rojas}

% RUNNING TITLE; SHORTEN THE TITLE IF NECESSARY
% IN CASE OF A ONE-PAGE CONTRIBUTION (POSTER),
% SQUEEZE AUTHORS AND TITLE IN THIS LINE (Author: Title ...)
\rohead{CNO trace in ring nebulae}

\begin{center}
% FULL TITLE HEADING
{\LARGE \bf Ring Nebulae: Tracers of the CNO Nucleosynthesis}\\
\medskip

% AUTHORS LIST
{\it\bf A. Mesa-Delgado$^1$, C. Esteban$^2$ \& J. Garc\'{\i}a-Rojas$^2$}\\

% AFFILIATIONS
{\it $^1$Instituto de Astrof\'{\i}sica, Facultad de F\'{\i}sica, Pontificia Universidad Cat\'olica de Chile, Av. Vicu\~na Mackenna 4860, 782-0436 Macul, Santiago, Chile}\\
{\it $^2$Instituto de Astrof\'{\i}sica de Canarias, E-38200 La Laguna, Tenerife, Spain}

% ABSTRACT
\begin{abstract}
Preliminary results are presented from spectroscopic data in the optical range of the Galactic ring nebulae NGC~6888, G$2.4+1.4$, RCW~58 and Sh2-308. Deep observations with long exposure times were carried out at the 6.5m Clay Telescope and at the 10.4m Gran Telescopio Canarias. In NGC~6888, recombination lines of C~{\sc ii}, O~{\sc ii} and N~{\sc ii} are detected with signal-to-noise ratios higher than 8. The chemical content of NGC~6888 is discussed within the chemical enrichment predicted by evolution models of massive stars. For all nebulae, a forthcoming work will content in-depth details about observations, analysis and final results (Esteban et al.~2015, in prep.).
\end{abstract}
\end{center}

% TEXT OF THE PAPER, TWO-COLUMN STYLE
\begin{multicols}{2}
%%%%%%%%%%%%%%
\section{Introduction}
Half a century has already passed since \cite{johnsonhogg65} pointed out the possible link between stellar evolution and the presence of fuzzy-filamentary, ring-like shells around Wolf-Rayet stars. Today, the origin and evolution of ring nebulae is certainly much better understood within the evolution of massive stars around and above 20~$M_\odot$. These circumstellar bubbles of ionized gas actually represent material of the surrounding medium swept up during the mass loss episodes experienced by their massive progenitors \citep[see e.g.][]{weaveretal77, freyeretal06, toalaarthur11}. A genuine phenomena that emerges from the interaction between the stellar wind and the environment, whose large-scale effects are observed in the chemical history of star-forming galaxies \citep[e.g.][]{lopezsanchezesteban10b}.

The classification of ring nebulae includes an interesting group: the ejected and wind-blown types as defined by \cite{chu81}. These ring nebulae show the presence of processed stellar material in their chemical composition. An enrichment process is behind these chemical traces. Through efficient dredge-up and mixing processes boosted by deep convective layers, nucleosynthetic products are transported from the stellar core to the surface. These products are mainly ejected during the post-main sequence, especially in the red supergiant (RSG), luminous blue variable (LBV) and Wolf-Rayet (WR) phases. This particular group of ring nebulae stands as powerful channels of information connected to the chemical history of previous evolutionary stages. Thus, their study is a unique approach to test our knowledge about massive stars, looking for a consistent picture between both stellar and nebular stories.  

Because of their inherent low surface brightness (F(H$\beta) \leq10^{-15}$ erg/s/cm$^2$/arcsec$^2$), the study of ring nebulae from their emission line spectra is indeed a challenging task. In the early 90s, Esteban and collaborators presented the first systematic study of the chemical composition of 11 Galactic ring nebulae \citep[e.g.][]{estebanetal92}. From deep spectroscopic data obtained from 2.5--4m telescope classes, these authors conducted a comprehensive analysis about the history of the stellar progenitors from the constraints given by the nebular story \citep{estebanetal93}. More recently, and mainly using a 3.5m telescope, \cite{stocketal11} have enhanced the sample of studied ring nebulae so far, even adding 3 ring nebulae of the Magellan Clouds. Consistently, the results show that ring nebulae containing stellar ejecta present overabundances of He and N, and a deficiency of O. A substantial fraction of O is transformed into N in the stellar interior via the activation of the ON chain and the He excess that remains until the activation of the $3\alpha$ reactions. This enrichment pattern is in agreement with the stellar nucleosynthesis of the H-burning through the CNO cycle during the main sequence of massive stars. 

For decades, a dramatic lack of reliable determinations of C abundances in any ring nebulae has hindered the full knowledge of the CNO cycle trace. The main spectral lines of C are either faint and challenging for detection as the recombination line C~{\sc ii}~4267~\AA, or either inaccessible from ground telescopes, requiring space facilities, as the doublet C~{\sc iii}]~1907+09~\AA. Recently, our research group reviewed the chemical composition of NGC~6888 with an improved accuracy (Mesa-Delgado~et~al.~2014; hereinafter, MD14) and reported the first detection of the C~{\sc ii} optical feature in this proto-typical ring nebula that is confirmed in this work. 
\begin{figure*}[!t]
\begin{center}
\includegraphics[scale=0.35]{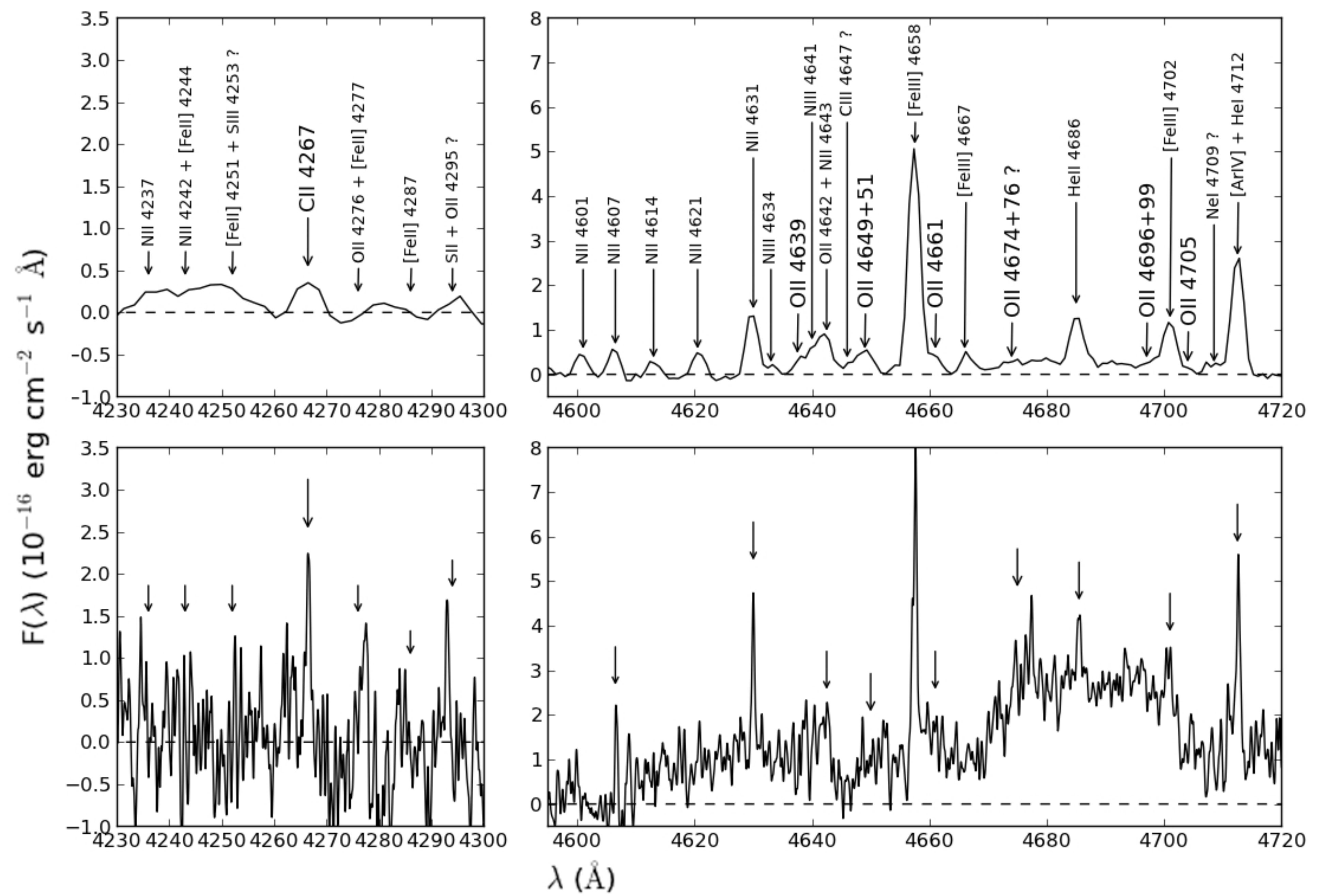}
\caption{Sections of NGC~6888 spectra. Top: detections of recombination lines of C~{\sc ii}, O~{\sc ii}, N~{\sc ii} and N~{\sc iii} from OSIRIS observations (Esteban et al.~2015, in prep.). Bottom: lower-limit detections of C~{\sc ii}, N~{\sc ii} and other features from HDS observations (MD14). The dashed lines represent the average continuum.
\label{mesadelgado_fig1}}
\end{center}
\end{figure*}

The present proceeding represents a preliminary {\it taste} of our renewed interest in reviewing the chemical composition of the brightest Galactic ring nebulae as revealed by deep spectroscopic observations with 6.5m and 10.4m telescopes. The final results will appear in a forthcoming work, including in-depth details about observations and analysis (Esteban et al.~2015, in prep.). 
%%%%%%%%%%%%%%
\section{Observations}
Making use of northern and southern facilities, we have collected the deepest spectroscopic data of the group of Galactic ring nebulae NGC~6888, G$2.4+1.4$, RCW~58 and Sh2-308. On the one hand, NGC~6888 and G$2.4+1.4$ were observed with the OSIRIS spectrograph at the 10.4m Gran Telescopio Canarias. Within its capabilities, this spectrograph provides a slit of $7.4'$ along the spatial direction, allowing us to simultaneously target multiple nebular areas and bright knots. On the other hand, single slit positions were observed on the brightest zones of G$2.4+1.4$, RCW~58 and Sh2-308 using the Magellan Echellette (MagE) spectrograph at the 6.5m Clay Telescope. In all cases, extractions of knot sizes, $3''$ to $8''$, were considered to maximize the signal-to-noise ratio of the final spectra. The two-observation campaigns were carried out with a slit width of $1''$, covering the optical range with a spectral resolution of about 3000. Total exposure times of 1.4 hours were used in the OSIRIS observations, while times between 1.2 and 2.5 hours were used in MagE.  

At first look, the outstanding quality and depth of these new observational data are well remarkable in the spectrum of NGC~6888. In Fig.~\ref{mesadelgado_fig1}, we compare the HDS spectra of NGC~6888 (MD14) with the recent OSIRIS observations. Both spectra represent similar extracted areas of about $1''\times4''$, located in positions on the bright arc to the northwest of the central star WR136, with similar integration times of 1.4 hours. The comparison is centered in the spectral ranges populated by recombination lines (RLs) of heavy-elements, and clearly reveals the superior result of the OSIRIS observations. The C~{\sc ii}~4267~\AA\ is well detected in this new dataset with a signal of about 8 over the continuum noise ($\sigma$). Almost featureless in the HDS observations, the O~{\sc ii} RLs appears blended among them and other lines. The eight transitions constituting the O~{\sc ii} multiplet 1 seem to be detected, where the brightest features are blended at 4649$+51$~\AA\ with a signal of about 17$\sigma$. N~{\sc ii} and N~{\sc iii} RLs are also detected, though their emission may be affected by continuum pumping \citep{escalantemorisset05}. The high-ionization feature of He~{\sc ii} 4686~\AA\ is also detected, making plausible the detection of the C~{\sc iii} line at 4647 \AA, the brightest feature of the transition 3s$^3$S--3p$^3$P$^0$.

\begin{figure*}[!t]
\begin{center}
\includegraphics[scale=0.42]{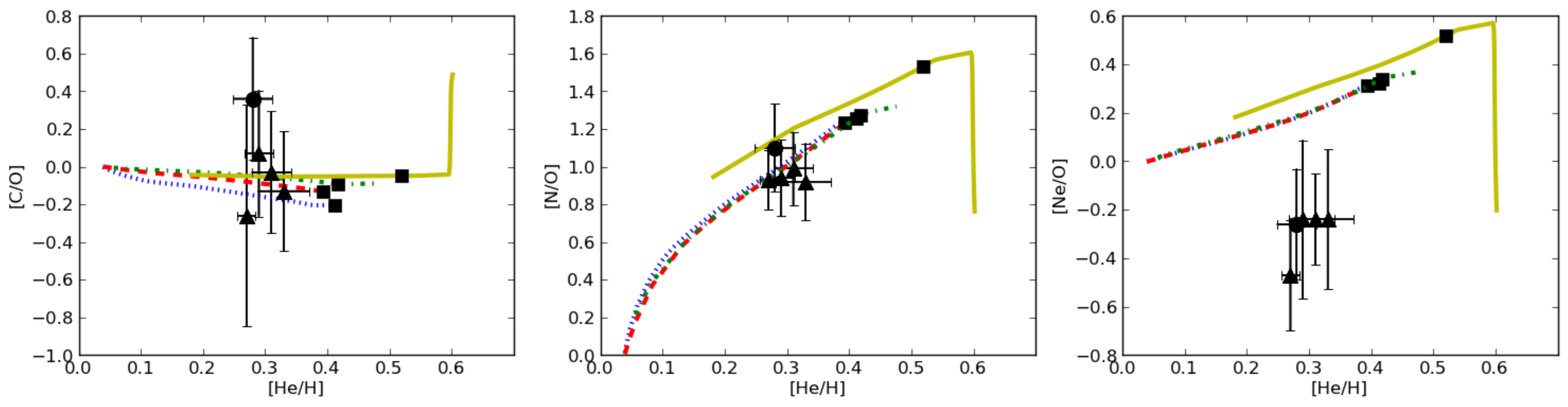}
\caption{Abundance ratios of C/O, N/O and Ne/O as a function of the He enrichment in NGC~6888. The zero points are referenced to the solar abundances as adopted by the Geneva models \citep{ekstrometal12}. Lines represent the enrichment predictions of non-rotational evolution models of massive stars since the RSG onset (see details in MD14). The masses are: 25~M$_\odot$ (blue-dotted line),  32~M$_\odot$ (red-dashed line), 40~M$_\odot$ (green-dotted-dashed line), and 50~M$_\odot$ (yellow-straight line). The square is the onset of the WR phase for each model. Observational data are presented with error bars: triangles represent OSIRIS observations and circles correspond to the MD14 study.
\label{mesadelgado_fig2}}
\end{center}
\end{figure*}

\section{Preliminary results}
The analysis of emission line spectra allows us to determine the chemical composition of these ring nebulae. Using the traditional methodology, He abundances are calculated from dereddened flux ratios of He~{\sc i} and He~{\sc ii} RLs, while chemical abundances of O, N and Ne are derived from collisionally excited lines (CELs). This chemical analysis is based on reliable detections ($>30\sigma$) of the temperature-sensitive auroral lines of [O~{\sc iii}] and [N~{\sc ii}], allowing us to calculate total abundances of He, N, and O with high accuracy ($\leq25$\%) in such faint objects as ring nebulae. The first results confirm the chemical enrichment previously reported in Sh2-308 and RCW~58, as well as the absence of enrichment in G$2.4+1.4$ \citep[see][]{estebanetal92}. 

In NGC~6888, both the new analysis of four OSIRIS extractions and the previous work by MD14 confirm the chemical enrichment by CNO cycle products. The final abundances derived by these studies are shown in Fig.~\ref{mesadelgado_fig2} together to the chemical enrichment traces predicted by the Geneva models for non-rotational evolution of stars with masses of 25, 32, 40 and 50~M$_\odot$ \citep{ekstrometal12}. These multi-positional analysis do not show any significant real abundance spread, favoring the presence of chemically homogeneous material within uncertainties. For instance, the optimal accuracy reached in the He abundances points out that, if chemical inhomogeneities may be presented in NGC~6888, they would produce chemical variations lower than 1\% across the nebula.   

For the first time in an ejecta nebula, the C/O abundance pattern is determined in a consistent manner from good detections of C~{\sc ii} and O~{\sc ii} RLs. In comparison with the OSIRIS results, we find that the C abundance derived by MD14 actually represents a higher limit (the highest C/O ratio in Fig.~\ref{mesadelgado_fig2}). Then, we estimate that the mean [C/O] ratio is essentially 0, consistent with the solar reference within the abundance errors. It is also consistent with the model predictions, where no C is produced in the CNO synthesis. In Fig.~\ref{mesadelgado_fig2}, the N enrichment becomes quite clear exploring the [N/O] ratio, which shows an average value of about 1. Certainly, the Ne/O ratio presents a puzzling behavior. An average value of $-0.3$ is observed in the [Ne/O] ratio, much lower than the predicted Ne enrichment by the models. As MD14 suggested, this behavior might be related to a higher efficiency of the NeNa cycle, higher than the assumed by current models, which also seems to correlate with the observations of Na deficiencies in the surface of yellow giant stars \citep{denissenkov05}. Finally, combining the abundance ratios of C/O, N/O and He/H and the characteristic lifetime of massive stars in the WR phase to develop such a bubble as NGC~6888 (see MD14), we constraint the mass of the stellar progenitor to be of about 40~M$_\odot$. 
{\small \paragraph{\small Acknowledgments: } A.M.D.~acknowledges support from the FONDECYT project 3140383. C.E.L.~and J.G.R.~acknowledge funding by MINECO under project AYA2011-22614.
}
%\bibliographystyle{aa} % style aa.bst
%\bibliography{biblio}

\end{multicols}

\end{contribution}

%%-------------------------------------------------------

\end{document}